\documentclass{article}
\usepackage{amssymb}

\input{tcilatex}
\usepackage{graphicx}

\begin{document}

\begin{center}
{\Large On a constructive characterization of a class of trees related to
pairs of disjoint matchings}

R.R. Kamalian$\dagger $, V. V. Mkrtchyan*$\dagger \ddagger $,

{\small *Department of Informatics and Applied Mathematics, Yerevan State
University, 0025, Armenia}

$\dagger ${\small Institute for Informatics and Automation Problems of
National Academy of Sciences of Armenia, 0014, Armenia}

{\small e-mail: rrkamalian@yahoo.com}

{\small vahanmkrtchyan2002@\{ysu.am, ipia.sci.am, yahoo.com,\}}

$\ddagger ${\small The author is supported by a grant of the Armenian
National Science and Educational Fund}
\end{center}

\bigskip

\begin{center}
\textbf{Abstract}
\end{center}

{\small For a graph consider the pairs of disjoint matchings which union
contains as many edges as possible, and define a parameter }$\alpha ${\small %
\ which eqauls the cardinality of the largest matching in those pairs. Also,
define }$\beta ${\small \ to be the cardinality of a maximum matching of the
graph.}

{\small We give a constructive characterization of trees which satisfy the }$%
\alpha =\beta ${\small \ equality. The proof of our main theorem is based on
a new decomposition algorithm obtained for trees.}


\bigskip

\begin{center}
\textbf{Introduction}\bigskip
\end{center}

Let $Z^{+}$ denote the set of non-negative integers. We consider finite,
undirected graphs without loops or multiple edges. Let $V(G)$ and $E(G)$
denote the sets of vertices and edges of a graph $G$, respectively.

If $v\in V(G)$ then let $d_{G}(v)$ denote the degree of a vertex $v$ in a
graph $G$. For a bridge $e=(v_{1},v_{2})$ of a connected graph $G$, let $%
G_{1},G_{2}$ be the connected components of $G-e$. Define the graphs $%
G_{1}e,G_{2}e$ as follows:

\begin{center}
$G_{1}e\equiv G\backslash (V(G_{2})\backslash \{v_{2}\}),$

$G_{2}e\equiv G\backslash (V(G_{1})\backslash \{v_{1}\}),$
\end{center}

where, without loss of generality, it is assumed, that $v_{i}\in
V(G_{i}),i=1,2$.

For a graph $G$, let $\beta (G)$ denote the cardinality of a maximum
matching of $G$. Define:

\begin{center}
$M(G)\equiv \{F:F$ is a maximum matching of $G\},$

$L(G)\equiv \max \{\beta (G\backslash F):F\in M(G)\},$

$M^{\prime }(G)\equiv \{F\in M(G):\beta (G\backslash F)=L(G)\}.$
\end{center}

Let us also define:

\begin{center}
$\lambda (G)\equiv \max \{\left\vert H\right\vert +\left\vert H^{\prime
}\right\vert :H,H^{\prime }$are matchings of $G$ with $H\cap H^{\prime
}=\varnothing \}$,

$M_{2}(G)\equiv \{(H,H^{\prime }):$ $\left\vert H\right\vert +\left\vert
H^{\prime }\right\vert =\lambda (G)$ and $H\cap H^{\prime }=\varnothing \}$,

$\alpha (G)\equiv \max \{\left\vert H\right\vert ,\left\vert H^{\prime
}\right\vert :$ $(H,H^{\prime })\in M_{2}(G)\}$,

$M_{2}^{\prime }(G)\equiv \{(H,H^{\prime }):(H,H^{\prime })\in
M_{2}(G),\left\vert H\right\vert =\alpha (G)\}.$
\end{center}

It is known that every graph $G$ contains a maximum $2$-matching that
includes a maximum matching of $G$ (see [7]). In contrast with the theory of 
$2$-matchings, in an arbitrary graph $G$ we cannot always guarantee the
existence of a "maximum" pair of disjoint matchings (i.e. pair of disjoint
matchings the union of which contains $\lambda (G)$ edges), which includes a
maximum matching. The following is the best we can do here: for every graph $%
G$ the following inequality is true [10]:

\begin{center}
$1\leq \frac{\beta (G)}{\alpha (G)}$ $\leq \frac{5}{4}$.
\end{center}

Let us also note that in her master thesis [11] Tserunyan gave an elegant
and very deep characterization of graphs which achieve the bound $\frac{5}{4}
$. Her theorem particularly implies that these graphs contain a spanning
subgraph every component of which is isomorph to the minimal graph that
satisfies the $\frac{\beta }{\alpha }=\frac{5}{4}$ equality.

In the light of this fact, the characterization of graphs which satisfy the $%
\alpha =\beta $ equality becomes a problem of notable importance. Moreover,
the problem is interesting not only because on its own but also because of
the equivalence :

\begin{center}
a graph $G$ satisfies the equality $\alpha (G)=\beta (G)$ if and only if $%
\lambda (G)=\beta (G)+L(G)$.
\end{center}

Though, the calculation of $\lambda (G)$ is $NP$-hard in general [4], the
Ford-Fulkerson algorithm for finding a maximum flow in a network implies
that it is indeed polynomial-time calculable for bipartite graphs. And, once
we are given a bipartite graph $G$ satisfying the equality $\alpha (G)=\beta
(G)$, we can calculate $L(G)$ easily.This is important, since $L(G)$ remains 
$NP$-hard calculable even for connected bipartite graphs $G$ with maximum
degree three [5]. Let us also note that there is a polynomial algorithm
which constructs a maximum matching $F$ of a tree $G$ such that $\beta
(G\backslash F)=L(G)$ (to be presented in [6]).

The aim of present paper is the characterization of trees that satisfy the $%
\alpha =\beta $ equality. An early result in this direction is given in [8]:
for every matching covered tree $G$ the equality $\alpha (G)=\beta (G)$
holds (a graph $G$ is referred to be matching covered if its every edge
belongs to a maximum matching of the graph [7, 9], complete characterization
of those trees can be found in [2,3]). The characterization given in the
paper is constructive, more specifically, we define four operations, with
the help of which we prove that a tree $G$ satisfies the equality $\alpha
=\beta $ if and only if it can be built from $K_{1}$ or $K_{2}$ (the trees
containing one or two vertices, respectively) by using these operations. Our
proof is based on a new decomposition algorithm obtained for the class of
trees.

Non-defined terms and concepts can be found in [1, 7, 12].

\bigskip

\begin{center}
\textbf{Some auxiliary results about }$\lambda (G),$\textbf{\ }$\alpha (G)$
and $L(G)$
\end{center}

\bigskip

\textbf{Lemma 1} Let $G$ be a graph, $v$ be a vertex with $d_{G}(v)=1$, and $%
e$ be the edge incident to it. Then

\begin{tabular}{ll}
1. [8] & There is $(H,H^{\prime })\in M_{2}^{\prime }(G)$, such that $e\in H$%
. \\ 
2. [6] & There is $F\in M^{\prime }(G)$, such that $e\in F$.%
\end{tabular}

\textbf{Lemma 2 }[8]. Let $G$ be a graph, $U=\{u_{0},...,u_{4}\}\subseteq
V(G)$ satisfying the conditions: $d_{G}(u_{0})=d_{G}(u_{4})=1,$ $%
d_{G}(u_{1})=d_{G}(u_{3})=2,$ $(u_{i-1},u_{i})\in E(G)$ for $i=1,2,3,4$ (fig
1). Then the following is true:

\begin{center}
$\lambda (G)=\lambda (G\backslash U)+4$, $\alpha (G)\geq 2+\alpha
(G\backslash U)$.

\FRAME{ftbpF}{5.348in}{3.0018in}{0in}{}{}{figure 1.jpg}{\raisebox{-3.0018in}{\includegraphics[height=3.0018in]{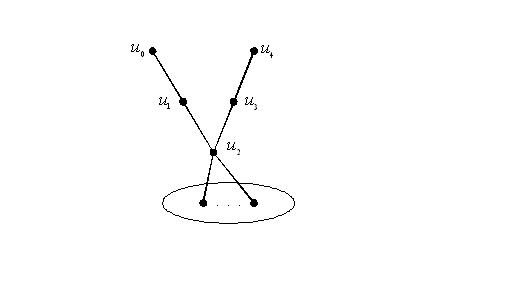}}}

Figure 1\bigskip
\end{center}

\textbf{Lemma 3}. Let $G$ be a graph and let $e\in E(G)$. Then

\begin{tabular}{ll}
(1) & $\lambda (G)\geq \lambda (G-e)$; \\ 
(2) & if $(H,H^{\prime })\in M_{2}(G)$ and $e\notin H\cup H^{\prime }$then $%
\lambda (G)=\lambda (G-e)$ and $\alpha (G)\geq \alpha (G-e)$; \\ 
(3) & if $(H,H^{\prime })\in M_{2}^{\prime }(G)$ and $e\notin H\cup
H^{\prime }$then $\alpha (G)=\alpha (G-e)$.%
\end{tabular}

\textbf{Lemma 4}. Let $G$ be a connected graph, $e$ be a bridge of $G$, and
let $G_{1}$, $G_{2}$ be the connected components of $G-e$. Then

\begin{tabular}{ll}
(1) & $\lambda (G)\geq \lambda (G_{1}e)+\lambda (G_{2}e)-1$; \\ 
(2) & if there is $(H,H^{\prime })\in M_{2}(G)$ with $e\in H\cup H^{\prime }$%
then $\lambda (G)=\lambda (G_{1}e)+\lambda (G_{2}e)-1$ and \\ 
& $\alpha (G)\geq \alpha (G_{1}e)+\alpha (G_{2}e)-1$; \\ 
(3) & if there is $(H,H^{\prime })\in M_{2}^{\prime }(G)$ with $e\in H$ then 
$\alpha (G)=\alpha (G_{1}e)+\alpha (G_{2}e)-1$.%
\end{tabular}

\textbf{Proof}. (1) Choose $(H_{1},H_{1}^{\prime })\in M_{2}^{\prime
}(G_{1}e)$, $(H_{2},H_{2}^{\prime })\in M_{2}^{\prime }(G_{2}e)$ with $e\in
H_{1},H_{2}$ ( (1) of lemma 1). Define:

\begin{center}
$H\equiv H_{1}\cup H_{2}$,

$H^{^{\prime }}\equiv H_{1}^{\prime }\cup H_{2}^{\prime }$.
\end{center}

Clearly, $H$ and $H^{\prime }$ are disjoint, and

\begin{center}
$\lambda (G)\geq \left\vert H\right\vert +\left\vert H^{\prime }\right\vert
=\left\vert H_{1}\right\vert +\left\vert H_{2}\right\vert -1+\left\vert
H_{1}^{\prime }\right\vert +\left\vert H_{2}^{\prime }\right\vert =\lambda
(G_{1}e)+\lambda (G_{2}e)-1$.
\end{center}

(2) Note that $(H\cap E(G_{1}e),H^{\prime }\cap E(G_{1}e))$ and $(H\cap
E(G_{2}e),H^{\prime }\cap E(G_{2}e))$\ are pairs of disjoint matchings in $%
G_{1}e$ and $G_{2}e$, respectively. Hence

\begin{center}
$\lambda (G)=\left\vert H\right\vert +\left\vert H^{\prime }\right\vert
=\left\vert H\cap E(G_{1}e)\right\vert +\left\vert H^{\prime }\cap
E(G_{1}e)\right\vert +\left\vert H\cap E(G_{2}e)\right\vert +$

$+\left\vert H^{\prime }\cap E(G_{2}e)\right\vert -1\leq \lambda
(G_{1}e)+\lambda (G_{2}e)-1$,
\end{center}

therefore

\begin{center}
$\lambda (G)=\lambda (G_{1}e)+\lambda (G_{2}e)-1$.
\end{center}

Note that this and lemma 1 imply that

\begin{center}
$\alpha (G)\geq \alpha (G_{1}e)+\alpha (G_{2}e)-1$.
\end{center}

(3) (2) implies that

\begin{center}
$(H\cap E(G_{1}e),H^{\prime }\cap E(G_{1}e))\in M_{2}(G_{1}e)$ and $(H\cap
E(G_{2}e),H^{\prime }\cap E(G_{2}e))\in M_{2}(G_{2}e)$,
\end{center}

hence

\begin{center}
$\alpha (G)=\left\vert H\right\vert =\left\vert H\cap E(G_{1}e)\right\vert
+\left\vert H\cap E(G_{2}e)\right\vert -1\leq \alpha (G_{1}e)+\alpha
(G_{2}e)-1$, or

$\alpha (G)=\alpha (G_{1}e)+\alpha (G_{2}e)-1$.
\end{center}

The proof of lemma 4 is completed.

\textbf{Lemma 5 }[6]. Let $G$ be a connected graph, $e$ be a bridge of $G$,
and let $G_{1}$, $G_{2}$ be the connected components of $G-e$. Then

\begin{center}
$L(G)=L(G_{1}e)+L(G_{2}e)$.
\end{center}

\bigskip

\begin{center}
\textbf{The main result}
\end{center}

\bigskip

\ In this section we introduce four elementary operations. They have the
property of preserving the equality $\beta =\alpha $, that is, if the graph
satisfies the equality then so does the graph obtained from original one by
the application of any of them. In the end of the section we prove that the
tree $G$ satisfying $\beta (G)=\alpha (G)$ can be built from $K_{1}$ or $%
K_{2}$ by using only these operations.

\bigskip

\ \textbf{Operation A}. Let $v_{1}$,...,$v_{k}$ ($k\geq 1$) be different
vertices of a graph $G$. Consider the graphs $G^{\prime }$ and $G^{\prime
\prime }$ obtained from $G$ in the following way (figure 2):

\begin{center}
\FRAME{ftbpF}{5.348in}{3.0018in}{0in}{}{}{figure 2.jpg}{\raisebox{-3.0018in}{\includegraphics[height=3.0018in]{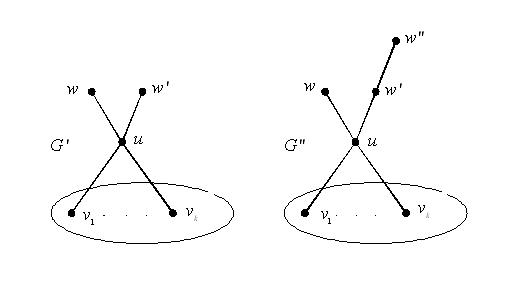}}}

Figure 2\bigskip
\end{center}

\ Since there are $(H_{1},H_{1}^{\prime })\in M_{2}^{\prime }(G^{\prime })$
and $(H_{2},H_{2}^{\prime })\in M_{2}^{\prime }(G^{\prime \prime })$ such
that $(u,v_{i})\notin H_{j}\cup H_{j}^{\prime },$ $1\leq i\leq k$ and $j=1,2 
$, we imply that (lemma 3)

\begin{center}
$%
\begin{array}{cc}
\alpha (G^{\prime })=1+\alpha (G), &  \\ 
& (\ast ) \\ 
\alpha (G^{\prime \prime })=2+\alpha (G). & 
\end{array}%
$
\end{center}

Note that the following equalities are also true [6]:

\begin{center}
$%
\begin{array}{cc}
\beta (G^{\prime })=1+\beta (G),L(G^{\prime })=1+L(G), &  \\ 
& (\ast \ast ) \\ 
\beta (G^{\prime \prime })=2+\beta (G),L(G^{\prime \prime })=1+L(G). & 
\end{array}%
$
\end{center}

Hence

\textbf{Lemma 6}. Either the graphs $G$, $G^{\prime }$, $G^{\prime \prime }$
satisfy the equality $\beta =\alpha $ or none of them does.\bigskip

Now, we proceed to the definitions of the three other operations. In
contrast with operation A, these ones are not always defined. This is the
main reason why the description of each operation is preceded by the
description of the cases when the operation is applicable.

\bigskip

\textbf{Operation B}.

\textbf{Definition 1}. A vertex $v$ of a graph $G$ is referred to be
applicable for the operation B if either $d_{G}(v)\leq 1$ or there is $%
U=\{u_{0},...,u_{4}\}\subseteq V(G)$ satisfying the conditions:

\begin{tabular}{ll}
(a) & $v=u_{2}$; \\ 
(b) & $(u_{i-1},u_{i})\in E(G)$ for $i=1,2,3,4$; \\ 
(c) & $d_{G}(u_{0})=1$, $d_{G}(u_{1})=d_{G}(u_{3})=2$ (figure 3)%
\end{tabular}

\begin{center}
\FRAME{ftbpF}{5.348in}{3.0018in}{0pt}{}{}{figure 3.jpg}{\raisebox{-3.0018in}{\includegraphics[height=3.0018in]{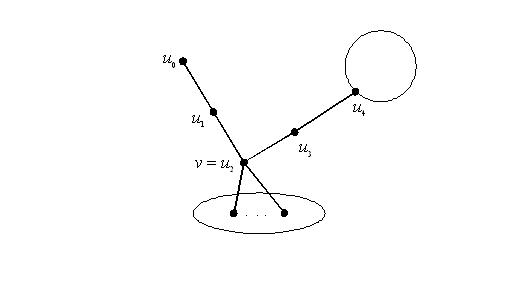}}}

Figure 3\bigskip
\end{center}

If $G$ is a graph, and $v$ is an applicable vertex for operation B, then $%
G^{\prime }$ (the result of operation B) is defined as follows (figure 4):

\begin{center}
\FRAME{ftbpF}{5.348in}{3.0018in}{0in}{}{}{figure 4.jpg}{\raisebox{-3.0018in}{\includegraphics[height=3.0018in]{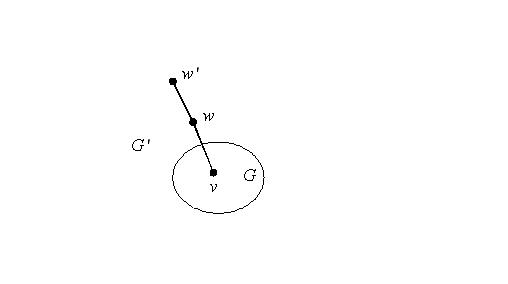}}}

\bigskip Figure 4
\end{center}

\textbf{Lemma 7}. $\beta (G^{\prime })=\alpha (G^{\prime })$ if and only if $%
\beta (G)=\alpha (G)$.

\textbf{Proof}. First of all note that $\beta (G^{\prime })=1+\beta (G)$.
The statement is true if $d_{G}(v)=0$. Assume that $d_{G}(v)=1$. Then

\begin{center}
$\lambda (G^{\prime })=2+\lambda (G)$,
\end{center}

and due to (1) of lemma 1 and (3) of lemma 4

\begin{center}
$\alpha (G^{\prime })=1+\alpha (G)$.
\end{center}

This shows that the statement of lemma 6 is true for the case of $d_{G}(v)=1$%
.

Therefore, we may assume that $d_{G}(v)\geq 2$. Since $v$ is applicable for
operation B, there is $U=\{u_{0},...,u_{4}\}\subseteq V(G)$ satisfying the
conditions (a), (b), (c) of definition 1. Let $\left\{ w,w^{\prime }\right\}
=V(G^{\prime })\backslash V(G)$ and $d_{G^{\prime }}(w)=2$, $d_{G^{\prime
}}(w^{\prime })=1$. Lemma 3 implies that to complete the proof it suffices
to show that there is $(H,H^{\prime })\in M_{2}(G^{\prime })$, such that $%
(w,v)\notin H\cup H^{\prime }$, or $(u_{1},v)\notin H\cup H^{\prime }$.

Choose any $(H,H^{\prime })\in M_{2}(G^{\prime })$, and assume that $%
\{(w,v),(u_{1},v)\}\subseteq H\cup H^{\prime }$. Without loss of generality,
we may assume that $\{(u_{0},u_{1}),(w,v)\}\subseteq H$ and $\{(w,w^{\prime
}),(u_{1},v)\}\subseteq H^{\prime }$. We claim that $(u_{3},u_{4})\in H$.
Suppose that $(u_{3},u_{4})\notin H$. Define:

\begin{center}
$\bar{H}\equiv (H\backslash \{(w,v)\})\cup \{(u_{3},v),(w^{\prime },w)\}$, $%
\bar{H}^{\prime }\equiv H^{\prime }\backslash \{(w^{\prime },w)\}$.
\end{center}

Note that

\begin{center}
$\left\vert \bar{H}\right\vert +\left\vert \bar{H}^{\prime }\right\vert
=\left\vert H\right\vert +\left\vert H^{\prime }\right\vert =\lambda
(G^{\prime })$ and $\left\vert \bar{H}\right\vert >\left\vert H\right\vert
=\alpha (G^{\prime })$,
\end{center}

which is impossible. Thus $(u_{3},u_{4})\in H$. Define:

\begin{center}
$H^{\prime \prime }\equiv (H^{\prime }\backslash \{(u_{1},v)\})\cup
\{(u_{2},u_{3})\}$.
\end{center}

Note that $(H,H^{\prime \prime })\in M_{2}(G^{\prime })$ and $%
\{(w,v),(u_{1},v)\}\nsubseteq H\cup H^{\prime \prime }$. The proof of lemma
7 is completed.

\bigskip

\textbf{Operation C}.

\textbf{Definition 2}. A vertex $v$ of a graph $G$ is referred to be
applicable for the operation C if either

(1) there is $U=\{u_{0},...,u_{6}\}\subseteq V(G)$ with

\qquad 
\begin{tabular}{ll}
(1a) & $v=u_{0}$; \\ 
(1b) & $d_{G}(u_{0})=d_{G}(u_{3})=d_{G}(u_{5})=1$, $%
d_{G}(u_{2})=d_{G}(u_{4})=2$, $d_{G}(u_{1})=4$, \\ 
& $(u_{i-1},u_{i})\in E(G)$ for $i=1,2,3,5$; $(u_{1},u_{4})\in E(G)$, $%
(u_{1},u_{6})\in E(G)$ (figure 5a); \\ 
(1c) & $\beta (He)=\beta (H)+1$, $L(He)=L(H)$, where $H\equiv G\backslash
(U\backslash \{u_{6}\})$ and $e=(u_{1},u_{6})$;%
\end{tabular}

or

(2) there is $U=\{u_{0},...,u_{4}\}\subseteq V(G)$ with

\qquad 
\begin{tabular}{ll}
(2a) & $v=u_{0}$; \\ 
(2b) & $d_{G}(u_{0})=d_{G}(u_{3})=1$, $d_{G}(u_{2})=2$, $d_{G}(u_{1})=3$, \\ 
& $(u_{i-1},u_{i})\in E(G)$ for $i=1,2,3$; $(u_{1},u_{4})\in E(G)$ (figure
5b); \\ 
(2c) & $\lambda (He)=\lambda (H)$, where $H\equiv G\backslash (U\backslash
u_{4})$, and $e=(u_{1},u_{4})$.%
\end{tabular}

\begin{center}
\FRAME{ftbpF}{5.348in}{3.096in}{0in}{}{}{figure 5.jpg}{\raisebox{-3.096in}{\includegraphics[height=3.096in]{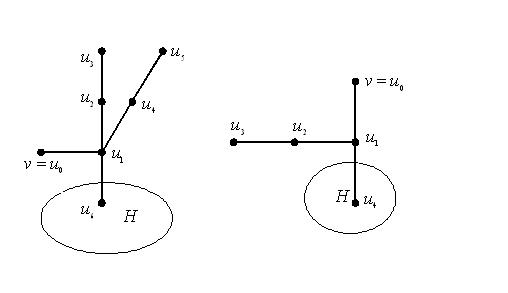}}}

Figure 5a \ \ \ \ \ \ \ \ \ \ \ \ \ \ \ \ \ \ \ \ \ \ \ \ \ \ \ \ \ \ \ \ \
\ \ \ \ \ \ \ \ \ \ \ Figure 5b\bigskip
\end{center}

If $G$ is a graph, and $v$ is an applicable vertex for operation C, then $%
G^{\prime }$ (the result of operation C) is defined as follows (figure 6):

\begin{center}
\FRAME{ftbpF}{5.348in}{3.0018in}{0in}{}{}{figure 6.jpg}{\raisebox{-3.0018in}{\includegraphics[height=3.0018in]{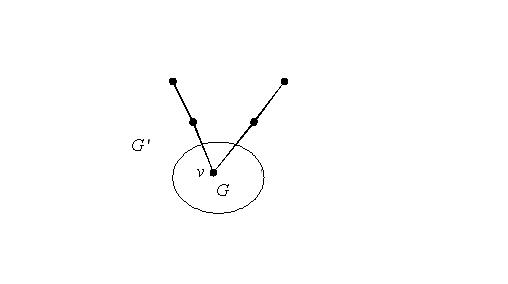}}}

Figure 6\bigskip
\end{center}

\textbf{Lemma 8}. If $\beta (G)=\alpha (G)$ then $\beta (G^{\prime })=\alpha
(G^{\prime })$.

\textbf{Proof}. Case 1: There is $U=\{u_{0},...,u_{6}\}\subseteq V(G)$
satisfying (1) of definition 2 (figure 5a).

Note that

\begin{center}
$\beta (G^{\prime })=2+\beta (G)=5+\beta (H)$ and due to (*)

$3+$ $\alpha (H)=\alpha (G)=\beta (G)=3+\beta (H)$, hence

$\alpha (H)=\beta (H)$, or $\lambda (H)=\beta (H)+L(H)$.
\end{center}

Let $\{g_{0},...,g_{3}\}=E(G^{\prime })\backslash E(G)$ (figure 7).

\begin{center}
\FRAME{ftbpF}{5.348in}{3.0018in}{0in}{}{}{figure 7.jpg}{\raisebox{-3.0018in}{\includegraphics[height=3.0018in]{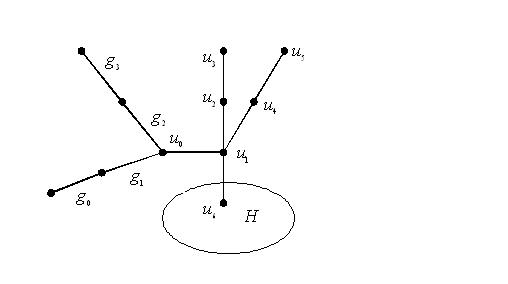}}}

Figure 7\bigskip
\end{center}

We claim that there is no $F$ $\in M^{\prime }(G^{\prime })$ containing the
edge $(u_{0},u_{1})$. Assume the contrary, and let $F\in M^{\prime
}(G^{\prime })$ contain the edge $(u_{0},u_{1})$.

Due to lemma 5

\begin{center}
$L(G^{\prime })=\beta (G^{\prime }\backslash F)=2+L(H)$.
\end{center}

Choose a maximum matching $F_{1}^{\prime }$ $\in M^{\prime }(He)$ (lemma 1).
Note that $e\in F_{1}^{\prime }$. Define:

\begin{center}
$F^{\prime }\equiv F_{1}^{\prime }\cup
\{g_{1},g_{3},(u_{2},u_{3}),(u_{4},u_{5})\}$.
\end{center}

Note that

\begin{center}
$\left\vert F^{\prime }\right\vert =4+\beta (He)=5+\beta (H)=\beta
(G^{\prime })$ and

$\beta (G^{\prime }\backslash F^{\prime })=3+L(He)=3+L(H)>L(G^{\prime })$
\end{center}

which is a contradiction.

This implies that there is $F^{\prime }$ $\in M^{\prime }(G^{\prime })$
containing $g_{1}$. Note that $e\in F^{\prime }$ (otherwise we would have an
augmenting path), therefore due to lemma 5

\begin{center}
$L(G^{\prime })=\beta (G^{\prime }\backslash F^{\prime })=3+L(He)$.
\end{center}

On the other hand, lemma 2 implies that

\begin{center}
$\lambda (G^{\prime })=8+\lambda (H)=8+\beta (H)+L(H)=8+\beta
(H)+L(He)=\beta (G^{\prime })+L(G^{\prime })$, hence

$\alpha (G^{\prime })=\beta (G^{\prime })$.
\end{center}

Case 2: There is $U=\{u_{0},...,u_{4}\}\subseteq V(G)$ satisfying (2) of
definition 2 (figure 5b).

Note that

\begin{center}
$\beta (G^{\prime })=2+\beta (G)=4+\beta (H)$ and due to (*)

$2+$ $\alpha (H)=\alpha (G)=\beta (G)=2+\beta (H)$, hence

$\alpha (H)=\beta (H)$, and

$\beta (He)+L(He)\leq \lambda (He)=\lambda (H)=\beta (H)+L(H)$.
\end{center}

Let $\{f_{0},...,f_{3}\}=E(G^{\prime })\backslash E(G)$ (figure 8).

\begin{center}
\FRAME{ftbpF}{5.348in}{3.0018in}{0in}{}{}{figure 8.jpg}{\raisebox{-3.0018in}{\includegraphics[height=3.0018in]{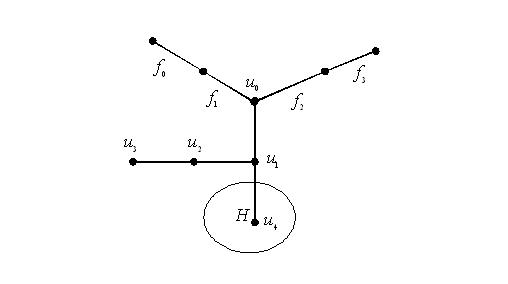}}}

Figure 8\bigskip
\end{center}

Let us show that there is $F^{\prime }\in M^{\prime }(G^{\prime })$
containing $(u_{0},u_{1})$. Take any $F$ $\in M^{\prime }(G^{\prime })$, and
assume that $(u_{0},u_{1})\notin F$. Note that

\begin{center}
$F\cap \{f_{1},f_{2}\}\neq \varnothing $ and $e\in F$ (otherwise we would
have an augmenting path).
\end{center}

Without loss of generality we may assume that $f_{1}\in F$. It is not hard
to see that

\begin{center}
$L(G^{\prime })=\beta (G^{\prime }\backslash F)=3+L(He)$ (lemma 5),

$4+\beta (H)=\beta (G^{\prime })=3+\beta (He)$, and therefore

$L(He)\leq L(H)-1$.
\end{center}

Let $F_{1}^{\prime }\in M^{\prime }(H)$. Define $F^{\prime }$ as follows:

\begin{center}
$F^{\prime }\equiv F_{1}^{\prime }\cup
\{f_{0},f_{3},(u_{0},u_{1}),(u_{2},u_{3})\}$.
\end{center}

Clearly

\begin{center}
$\left\vert F^{\prime }\right\vert =4+\beta (H)=\beta (G^{\prime })$ hence $%
F^{\prime }\in M(G^{\prime })$, and

$\beta (G^{\prime }\backslash F^{\prime })=2+L(H)\geq 3+L(He)=L(G^{\prime })$%
,
\end{center}

hence $F^{\prime }\in M^{\prime }(G^{\prime })$ and $(u_{0},u_{1})\in
F^{\prime }$. Lemma 5 and (**) imply that

\begin{center}
$L(G^{\prime })=2+L(H)$.
\end{center}

Lemmata 2,4 imply that

\begin{center}
$\lambda (G^{\prime })=6+\lambda (He)=6+\lambda (H)=6+\beta (H)+L(H)=\beta
(G^{\prime })+L(G^{\prime })$, hence

$\alpha (G^{\prime })=\beta (G^{\prime })$.
\end{center}

The proof of lemma 8 is completed.

\bigskip

\textbf{Operation D}.

\textbf{Definition 3}. A vertex $v$ of a graph $G$ is referred to be
applicable for the operation D if either

(1) there is $U=\{u_{0},u_{1},u_{2}\}\subseteq V(G)$ with

\qquad 
\begin{tabular}{ll}
(1a) & $v=u_{1}$; \\ 
(1b) & $d_{G}(u_{0})=1$, $d_{G}(u_{1})=2$, $(u_{i-1},u_{i})\in E(G)$ for $%
i=1,2$ (figure 9a);%
\end{tabular}

or

(2) there is $U=\{u_{0},...,u_{5}\}\subseteq V(G)$ with

\qquad 
\begin{tabular}{ll}
(2a) & $v=u_{5}$; \\ 
(2b) & $d_{G}(u_{0})=d_{G}(u_{4})=1$, $d_{G}(u_{1})=d_{G}(u_{3})=2$, $%
d_{G}(u_{2})=3$, \\ 
& $(u_{i-1},u_{i})\in E(G)$ for $i=1,2,3,4$; $(u_{2},u_{5})\in E(G)$ (figure
9b);%
\end{tabular}

or

(3) there is $U=\{u_{0},...,u_{3}\}\subseteq V(G)$ with

\qquad 
\begin{tabular}{ll}
(3a) & $v=u_{2}$; \\ 
(3b) & $d_{G}(u_{0})=1$, $d_{G}(u_{1})=d_{G}(u_{2})=2$, $(u_{i-1},u_{i})\in
E(G)$ for $i=1,2,3$ (figure 9c); \\ 
(3c) & $\beta (He)=\beta (H)+1$, $L(He)=L(H)$, where $H\equiv G\backslash
(U\backslash \{u_{3}\})$, and $e=(u_{2},u_{3})$.%
\end{tabular}

\begin{center}
\FRAME{ftbpF}{5.348in}{3.0649in}{0in}{}{}{figure 9.jpg}{\raisebox{-3.0649in}{\includegraphics[height=3.0649in]{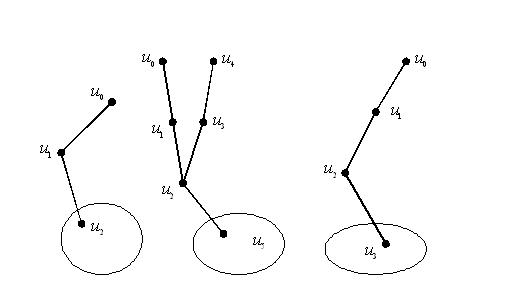}}}

Figure 9a \qquad \qquad \qquad Figure 9b \qquad \qquad \qquad Figure
9c\bigskip
\end{center}

If $G$ is a graph, and $v$ is an applicable vertex for operation D, then $%
G^{\prime }$ (the result of operation D) is defined as follows (figure 10):

\begin{center}
\FRAME{ftbpF}{5.348in}{3.0018in}{0in}{}{}{figure 10.jpg}{\raisebox{-3.0018in}{\includegraphics[height=3.0018in]{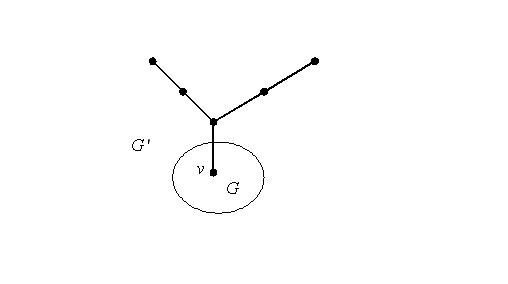}}}

Figure 10\bigskip
\end{center}

\textbf{Lemma 9}. If $\beta (G)=\alpha (G)$ then $\beta (G^{\prime })=\alpha
(G^{\prime })$.

\textbf{Proof}. Case 1: There is $U=\{u_{0},u_{1},u_{2}\}\subseteq V(G)$
satisfying (1) of definition 3 (figure 9a).

Note that lemma 2 implies that

\begin{center}
$\beta (G^{\prime })=2+\beta (G)$,

$\lambda (G^{\prime })=4+\lambda (G)$, therefore

$\alpha (G^{\prime })\geq 2+\alpha (G)=2+\beta (G)=\beta (G^{\prime })$, or

$\alpha (G^{\prime })=\beta (G^{\prime })$.
\end{center}

Case 2: There is $U=\{u_{0},...,u_{5}\}\subseteq V(G)$ satisfying (2) of
definition 3 (figure 9b).

From lemma 2 we have

\begin{center}
$\beta (G^{\prime })=2+\beta (G)$,

$\lambda (G^{\prime })=4+\lambda (G)$, therefore

$\alpha (G^{\prime })\geq 2+\alpha (G)=2+\beta (G)=\beta (G^{\prime })$, or

$\alpha (G^{\prime })=\beta (G^{\prime })$.
\end{center}

Case 3: There is $U=\{u_{0},...,u_{3}\}\subseteq V(G)$ satisfying (3) of
definition 3 (figure 9c).

Note that

\begin{center}
$\beta (G^{\prime })=4+\beta (H)$,

$\beta (G)=1+\beta (He)$,

$\alpha (G)=1+\alpha (He)$ (lemma 4), hence

$\alpha (He)=\beta (He)$, or $\lambda (He)=\beta (He)+L(He)$
\end{center}

Let $\{g_{0},...,g_{4}\}=E(G^{\prime })\backslash E(G)$ (figure 11).

\begin{center}
\FRAME{ftbpF}{5.348in}{3.0649in}{0in}{}{}{figure 11.jpg}{\raisebox{-3.0649in}{\includegraphics[height=3.0649in]{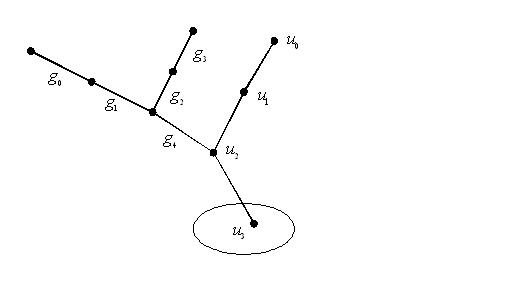}}}

Figure 11\bigskip
\end{center}

We claim that there is no $F^{\prime }\in M^{\prime }(G^{\prime })$
containing the edge $g_{4}$. On the opposite assumption, consider $F^{\prime
}\in M^{\prime }(G^{\prime })$ with $g_{4}\in F^{\prime }$. Note that

\begin{center}
$\{g_{0},g_{3},(u_{0},u_{1})\}\subseteq F^{\prime }$ and $L(G^{\prime
})=\beta (G^{\prime }\backslash F^{\prime })=2+L(H)$ (lemma 5).
\end{center}

Let $F_{1}\in M^{\prime }(He)$. Note that $e\in F_{1}$. Define $F$ as
follows:

\begin{center}
$F\equiv F_{1}\cup \{g_{1},g_{3},(u_{0},u_{1})\}$.
\end{center}

Clearly

\begin{center}
$\left\vert F\right\vert =3+\beta (He)=4+\beta (H)=\beta (G^{\prime })$ and

$\beta (G^{\prime }\backslash F)=3+L(H)>2+L(H)=L(G^{\prime })$,
\end{center}

which is a contradiction. This implies that there is $F$ $\in M^{\prime
}(G^{\prime })$ containing $g_{1}$. Note that as $e\in F$ (otherwise we
would have an augmenting path), we imply that

\begin{center}
$L(G^{\prime })=\beta (G^{\prime }\backslash F)=3+L(He)=3+L(H)$ (lemma 5).
\end{center}

On the other hand, lemma 2 implies that (see the definition of operation B)

\begin{center}
$\lambda (G^{\prime })=4+\lambda (G)=6+\lambda (He)=6+\beta
(He)+L(He)=7+\beta (H)+L(H)=\beta (G^{\prime })+L(G^{\prime })$,
\end{center}

hence

\begin{center}
$\alpha (G^{\prime })=\beta (G^{\prime })$.
\end{center}

The proof of lemma 9 is completed.

\bigskip

\textbf{Theorem}. A tree $G$ satisfies the equality $\beta (G)=\alpha (G)$
if and only if it is either $K_{1}$ or $K_{2}$, or can be obtained from them
by the application of the operations A, B, C or D.

\textbf{Proof}. Note that\ $K_{1}$ and $K_{2}$ satisfy the equality $\beta
=\alpha $, and lemmata 6,7,8,9 imply that the operations A, B, C or D
preserve this property, that is, whatever tree $G$ we build from $K_{1}$ or $%
K_{2}$ by these operations we will always have $\beta (G)=\alpha (G)$.

Let us show that the converse is also true, i.e. every tree $G$ satisfying $%
\beta (G)=\alpha (G)$ can be built from $K_{1}$ or $K_{2}$ by A, B, C or D.

The proof is on induction. Clearly, the statement is true if $\left\vert
E(G)\right\vert \leq 1$. Assume that the statement is true for all trees $%
G^{\prime }$ which satisfy the equality $\beta (G^{\prime })=\alpha
(G^{\prime })$ and $\left\vert E(G^{\prime })\right\vert <\left\vert
E(G)\right\vert $, and let us show that it also holds for the tree $G$
satisfying $\beta (G)=\alpha (G)$.

First of all note that we may always assume that there is no $%
U=\{u_{0},u_{1},u_{2}\}\subseteq V(G)$ with $d_{G}(u_{0})=1$, $%
d_{G}(u_{1})=d_{G}(u_{2})=2$, $(u_{i-1},u_{i})\in E(G)$ for $i=1,2$. On the
opposite assumption, consider the set $U$ comprised of vertices $%
u_{0},u_{1},u_{2}$ satisfying these conditions. Set:

\begin{center}
$G^{\prime }\equiv G\backslash \{u_{0},u_{1}\}$.
\end{center}

The definition of operation B and lemma 4 imply that $\beta (G^{\prime
})=\alpha (G^{\prime })$. The induction hypothesis implies that $G^{\prime }$
can be built from $K_{1}$ or $K_{2}$ by A, B, C or D, and since $G$ can be
built from $G^{\prime }$ by operation B, we are done.

Now let us show that we may also assume that there is no $%
U=\{u_{0},...,u_{6}\}\subseteq V(G)$ with $(u_{i-1},u_{i})\in E(G)$ for $%
i=1,2,3,4,6$; $(u_{2},u_{5})\in E(G)$, $%
d_{G}(u_{0})=d_{G}(u_{4})=d_{G}(u_{6})=1$, $%
d_{G}(u_{1})=d_{G}(u_{3})=d_{G}(u_{5})=2$. If $U=\{u_{0},...,u_{6}\}$ is
such a set, then set:

\begin{center}
$G^{\prime }\equiv G\backslash \{u_{0},u_{1}\}$.
\end{center}

The definition of operation B and lemma 7 imply that $\beta (G^{\prime
})=\alpha (G^{\prime })$ and therefore due to induction hypothesis, $%
G^{\prime }$ can be built from $K_{1}$ or $K_{2}$ by A, B, C or D. As $%
u_{2}\in V(G\backslash \{u_{0},u_{1}\})$ is applicable for B and $G$ is
built from $G^{\prime }$ by applying B, we conclude that $G$ can be built
from $K_{1}$ or $K_{2}$ by A, B, C or D.

Define:

\begin{center}
$V_{G}(0)\equiv \{v\in V(G):d_{G}(v)=1\}$,
\end{center}

and for $i\geq 1$ let

\begin{center}
$V_{G}(i)\equiv \{v\in V(G):d_{H}(v)=1,$ where $H\equiv G\backslash
(\tbigcup\limits_{j=0}^{i-1}V_{G}(j))\}$.
\end{center}

Consider a mapping $k_{G}:V(G)\rightarrow Z^{+}$ defined as:

\begin{center}
for $v\in V(G)$ $\ v\in V_{G}(k_{G}(v))$.
\end{center}

Note that for each vertex $v$ there is at most one vertex $v^{\prime }$ with 
$(v,v^{\prime })\in E(G)$\ and $k_{G}(v^{\prime })>k_{G}(v)$.

Since $G$ is not a path, we imply that it contains a vertex of degree at
least three. Now, choose a vertex $v\in V(G)$ satisfying the conditions:

\begin{center}
$d_{G}(v)\geq 3$ and $k_{G}(v)\rightarrow \min $.
\end{center}

Note that the choice of $v$ implies that there are paths $P_{1},...,P_{r}$ ($%
r\geq 2$) of $G$ satisfying the conditions:

\begin{center}
for every $w\in V(P_{i})$ $1\leq i\leq r$, $w\neq v$, $d_{G}(w)\leq 2$ and $%
k_{G}(w)<k_{G}(v)$;

$d_{G}(v)=r+1$ (figure 12).

\FRAME{ftbpF}{5.348in}{3.0018in}{0pt}{}{}{figure 12.jpg}{\raisebox{-3.0018in}{\includegraphics[height=3.0018in]{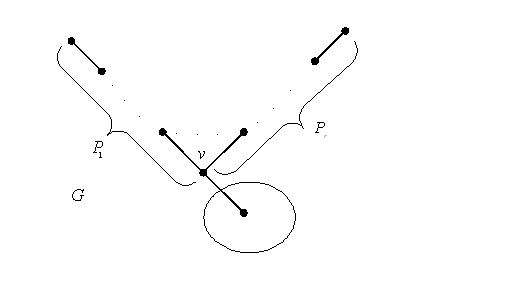}}}

Figure 12\bigskip
\end{center}

We claim that without loss of generality we may assume that $r=2$ and $%
P_{1},P_{2}$ are of length two for every vertex $v\in V(G)$ satisfying the
conditions $d_{G}(v)\geq 3$ and $k_{G}(v)\rightarrow \min $.

Note that every path from $P_{1},...,P_{r}$ is of length at most two. Now,
let us show that paths $P_{1},...,P_{r}$ may be assumed to have lengths
equal to two. Let $P_{1}$ have a length equal to one, and let $%
V(P_{1})=\{u,v\}$. Consider the trees $G_{1},...,G_{r-1}$ - the connected
components of $G\backslash (V(P_{1})\cup V(P_{2}))$. Note that (see
operation A)

\begin{center}
$\beta (G)-\alpha (G)=\tsum\limits_{i=1}^{r-1}(\beta (G_{i})-\alpha (G_{i}))$%
,
\end{center}

and since $\beta (G)=\alpha (G)$ we imply that $\beta (G_{i})=\alpha (G_{i})$%
, $i=1,...,r-1$. Due to hypothesis of induction we conclude that $G_{i}$, $%
1\leq i\leq r-1$, can be built from $K_{1}$ or $K_{2}$ by A, B, C or D. Note
that since $G$ is built from $G\backslash (V(P_{1})\cup V(P_{2}))$ by
operation A, we are done. This shows that the lengths of paths $%
P_{1},...,P_{r}$ may be assumed to be equal to two, and therefore we may
also assume that $r=2$ for every vertex $v\in V(G)$ satisfying the
conditions: $d_{G}(v)\geq 3$ and $k_{G}(v)\rightarrow \min $.

As $G$ is not a path and $\beta (G)=\alpha (G)$ we imply that for every
vertex $v\in V(G)$ with $d_{G}(v)\geq 3$ and $k_{G}(v)\rightarrow \min $
there is a unique $v^{\prime }\in V(G)$ such that $(v,v^{\prime })\in E(G)$
and $k_{G}(v)<k_{G}(v^{\prime })$.

Now, choose a vertex $v\in V(G)$ satisfying the conditions:

\begin{center}
$d_{G}(v)\geq 3$, $k_{G}(v)\rightarrow \min $ and $k_{G}(v^{\prime
})\rightarrow \min $,
\end{center}

where $v^{\prime }$ is the abovementioned vertex corresponding to $v$.

Note that the choice of $v$ implies that $d_{G}(v^{\prime })\geq 2$ and $%
k_{G}(v^{\prime })=k_{G}(v)+1$. Let us show that we may also assume that $%
d_{G}(v^{\prime })\geq 3$. Suppose that $d_{G}(v^{\prime })=2$, and let $%
u_{1},...,u_{4},v^{\prime \prime }$ be vertices shown in the figure below:

\begin{center}
\FRAME{ftbpF}{5.348in}{3.0018in}{0in}{}{}{figure 13.jpg}{\raisebox{-3.0018in}{\includegraphics[height=3.0018in]{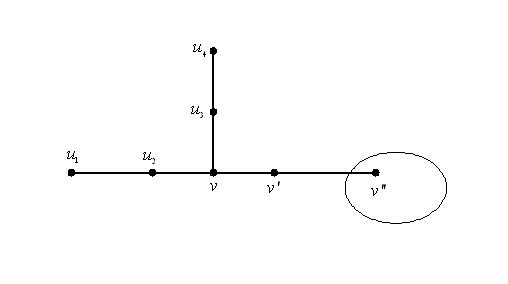}}}

Figure 13\bigskip
\end{center}

Let us show that there is $(H,H^{\prime })\in M_{2}^{\prime }(G)$ such that

\begin{center}
$\left\{ (u_{2},v),(u_{3},v)\right\} \nsubseteq H\cup H^{\prime }$.
\end{center}

Choose $(H,H^{\prime })\in M_{2}^{\prime }(G)$ and, without loss of
generality, assume that $(u_{2},v)\in H$, $(u_{3},v)\in H^{\prime }$.
Define: $H_{1}$ and $H_{1}^{\prime }$ as follows:

\begin{center}
$H_{1}\equiv H$, $H_{1}^{\prime }\equiv (H^{\prime }\backslash
\{(u_{3},v)\})\cup \{(v,v^{\prime })\}$ if $(v^{\prime },v^{\prime \prime
})\in H$,

$H_{1}\equiv (H\backslash \{(u_{2},v)\})\cup \{(v,v^{\prime })\}$, $%
H_{1}^{\prime }\equiv H^{\prime }$ if $(v^{\prime },v^{\prime \prime })\in
H^{\prime }$.
\end{center}

Note that $(H_{1},H_{1}^{\prime })\in M_{2}^{\prime }(G)$ and $\left\{
(u_{2},v),(u_{3},v)\right\} \nsubseteq H_{1}\cup H_{1}^{\prime }$.

It is not hard to see that this implies that there is $(H_{2},H_{2}^{\prime
})\in M_{2}^{\prime }(G)$ such that $(u_{2},v)\notin H_{2}\cup H_{2}^{\prime
}$. lemma 3 implies that

\begin{center}
$\alpha (G\backslash \{u_{1},u_{2}\})=\alpha (G\backslash
\{(u_{2},v)\})-1=\alpha (G)-1=\beta (G)-1=\beta (G\backslash
\{u_{1},u_{2}\}) $,
\end{center}

hence the tree $G\backslash \{u_{1},u_{2}\}$ also satisfies the $\beta
=\alpha $ equality. Due to hypothesis of induction $G\backslash
\{u_{1},u_{2}\}$ can be built from $K_{1}$ or $K_{2}$ by A, B, C or D. Note
that $G$ is obtained from $G\backslash \{u_{1},u_{2}\}$ by operation B since
the vertex $v$ is applicable for it. This shows that $G$ can also be built
from $K_{1}$ or $K_{2}$ by A, B, C or D.

Thus, we may assume that $d_{G}(v^{\prime })\geq 3$. Let us show that we may
also assume that $v^{\prime }$ is not adjacent to a vertex $u$ with $%
d_{G}(u)=1$. On the opposite assumption, consider a vertex $u$ satisfying
conditions: $d_{G}(u)=1$ and $(u,v^{\prime })\in E(G)$. Let $u_{1},...,u_{4}$
be vertices shown in the figure below:

\begin{center}
\FRAME{ftbpF}{5.348in}{3.0018in}{0in}{}{}{figure 14.jpg}{\raisebox{-3.0018in}{\includegraphics[height=3.0018in]{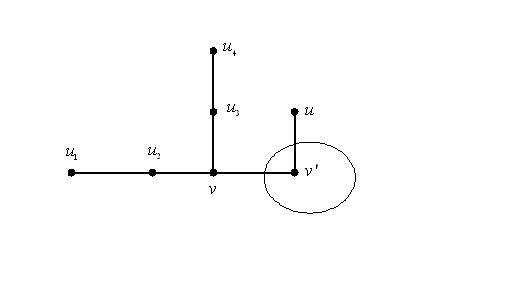}}}

Figure 14\bigskip
\end{center}

We claim that there is $(H,H^{\prime })\in M_{2}^{\prime }(G)$ with $%
(v,v^{\prime })\notin H\cup H^{\prime }$. Take any $(H,H^{\prime })\in
M_{2}^{\prime }(G)$ with $(u,v^{\prime })\in H$ (lemma 1), and suppose that $%
(v,v^{\prime })\in H^{\prime }$. Note that one of the edges $(u_{2},v)$ and $%
(u_{3},v)$ does not belong to $H\cup H^{\prime }$. Assume that $%
(u_{2},v)\notin H\cup H^{\prime }$. Since $\left\vert H\right\vert =\alpha
(G)$ we have $\left( u_{1},u_{2}\right) \in H$. Define:

\begin{center}
$H^{\prime \prime }\equiv (H^{\prime }\backslash \{(v,v^{\prime })\})\cup
\{(u_{2},v)\}$.
\end{center}

Note that $(H,H^{\prime \prime })\in M_{2}^{\prime }(G)$ and $(v,v^{\prime
})\notin H\cup H^{\prime \prime }$. This and lemma 3 imply that

\begin{center}
$\alpha (G\backslash \{v,u_{1},...,u_{4}\})=\alpha (G\backslash
\{(v,v^{\prime })\})-2=\alpha (G)-2=\beta (G)-2=\beta (G\backslash
\{v,u_{1},...,u_{4}\})$,
\end{center}

hence the tree $G\backslash \{v,u_{1},...,u_{4}\}$ also satisfies the $\beta
=\alpha $ equality. Due to hypothesis of induction $G\backslash
\{v,u_{1},...,u_{4}\}$ can be built from $K_{1}$ or $K_{2}$ by A, B, C or D.
Note that $G$ is obtained from $G\backslash \{v,u_{1},...,u_{4}\}$ by
operation D since the vertex $v$ is applicable for it. This shows that $G$
can also be built from $K_{1}$ or $K_{2}$ by A, B, C or D.

Thus, we may assume that $v^{\prime }$ is not adjacent to a vertex $u$ with $%
d_{G}(u)=1$. Now, we claim that we may assume that there is no a vertex $%
\bar{v}\neq v$ such that

\begin{center}
$d_{G}(\bar{v})\geq 3$, $k_{G}(\bar{v})=$ $k_{G}(v)\rightarrow \min $ and $(%
\bar{v},v^{\prime })\in E(G)$.
\end{center}

On the opposite assumption, consider a vertex $\bar{v}$ satisfying these
conditions, and let $u_{1},...,u_{8}$ be vertices shown in the figure below:

\begin{center}
\FRAME{ftbpF}{5.348in}{3.0018in}{0in}{}{}{figure 15.jpg}{\raisebox{-3.0018in}{\includegraphics[height=3.0018in]{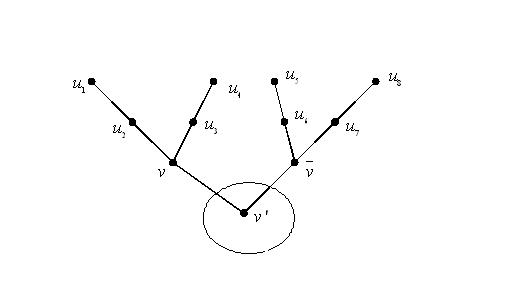}}}

Figure 15\bigskip
\end{center}

We claim that there is $(H,H^{\prime })\in M_{2}^{\prime }(G)$ with $%
(v,v^{\prime })\notin H\cup H^{\prime }$. Take any $(H,H^{\prime })\in
M_{2}^{\prime }(G)$.

Case 1: $(v,v^{\prime })\in H^{\prime }$. Note that one of the edges $%
(u_{2},v)$ and $(u_{3},v)$ does not belong to $H\cup H^{\prime }$. Assume
that $(u_{2},v)\notin H\cup H^{\prime }$. Since $\left\vert H\right\vert
=\alpha (G)$ we have $\left( u_{1},u_{2}\right) \in H$. Define:

\begin{center}
$H^{\prime \prime }\equiv (H^{\prime }\backslash \{(v,v^{\prime })\})\cup
\{(u_{2},v)\}$.
\end{center}

Note that $(H,H^{\prime \prime })\in M_{2}^{\prime }(G)$ and $(v,v^{\prime
})\notin H\cup H^{\prime \prime }$.

Case 2: $(v,v^{\prime })\in H$. Define $\bar{H}$, $\bar{H}^{\prime }$ as
follows:

\begin{center}
$\bar{H}\equiv (H\cap E(G\backslash \{v,v^{\prime },\bar{v}%
,u_{1},...,u_{8}\}))\cup \{(v^{\prime },\bar{v}%
),(u_{2},v),(u_{3},u_{4}),(u_{5},u_{6}),(u_{7},u_{8})\}$,

$\bar{H}^{\prime }\equiv (H^{\prime }\cap E(G\backslash \{v,v^{\prime },\bar{%
v},u_{1},...,u_{8}\}))\cup \{(u_{7},\bar{v}),(u_{1},u_{2}),(u_{3},v)\}$.
\end{center}

Clearly, $(\bar{H},\bar{H}^{\prime })\in M_{2}^{\prime }(G)$ and $%
(v,v^{\prime })\notin \bar{H}\cup \bar{H}^{\prime }$.

This and lemma 3 imply that

\begin{center}
$\alpha (G\backslash \{v,u_{1},...,u_{4}\})=\alpha (G\backslash
\{(v,v^{\prime })\})-2=\alpha (G)-2=\beta (G)-2=\beta (G\backslash
\{v,u_{1},...,u_{4}\})$,
\end{center}

hence the tree $G\backslash \{v,u_{1},...,u_{4}\}$ also satisfies the $\beta
=\alpha $ equality. Due to hypothesis of induction $G\backslash
\{v,u_{1},...,u_{4}\}$ can be built from $K_{1}$ or $K_{2}$ by A, B, C or D.
Note that $G$ is obtained from $G\backslash \{v,u_{1},...,u_{4}\}$ by
operation D since the vertex $v$ is applicable for it. This shows that $G$
can also be built from $K_{1}$ or $K_{2}$ by A, B, C or D.

Thus, we may assume that $v^{\prime }$ is not adjacent to another vertex $%
\bar{v}$ satisfying the conditions:

\begin{center}
$d_{G}(\bar{v})\geq 3$, $k_{G}(\bar{v})=$ $k_{G}(v)\rightarrow \min $.
\end{center}

It is not hard to see that there are paths $P_{1},...,P_{r}$ ($1\leq r\leq 2$%
) starting from the vertex $v^{\prime }$ and satisfying the conditions:

\begin{center}
for every $w\in V(P_{i})$ $1\leq i\leq r$, $w\neq v^{\prime }$, $%
d_{G}(w)\leq 2$ and $k_{G}(w)<k_{G}(v^{\prime })$.
\end{center}

Now, we will consider the remaining two cases:

Case 1: $r=2$. Let $v^{\prime \prime },u_{1},...,u_{8}$ be vertices shown in
the figure below:

\begin{center}
\FRAME{ftbpF}{5.348in}{3.0018in}{0in}{}{}{figure 16.jpg}{\raisebox{-3.0018in}{\includegraphics[height=3.0018in]{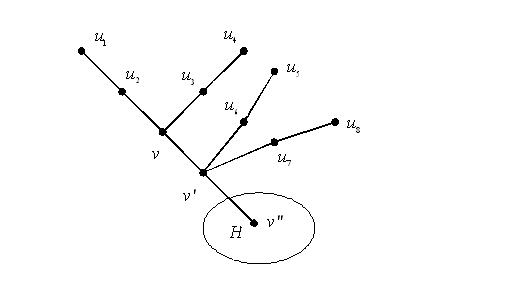}}}

Figure 16\bigskip
\end{center}

Assume:

\begin{center}
$H\equiv G\backslash \{v,v^{\prime },u_{1},...,u_{8}\}$, $e\equiv (v^{\prime
},v^{\prime \prime })$.
\end{center}

We claim that there is no $F\in M^{\prime }(G)$ containing the edge $%
(v,v^{\prime })$. Suppose there is. Note that

\begin{center}
$\beta (G)=5+\beta (H)$,

$L(G)=\beta (G\backslash F)=2+L(H)$ (lemma 5),
\end{center}

and since $\alpha (G)=\beta (G)$, we have

\begin{center}
$\lambda (G)=\beta (G)+L(G)=7+\lambda (H)$,
\end{center}

contradicting lemma 2 which imples that

\begin{center}
$\lambda (G)=8+\lambda (H)$.
\end{center}

This immediately implies that $\beta (He)=1+\beta (H)$ and, consequently, $%
L(He)\leq L(H)$. Let us show that $\ L(He)=L(H)$. Suppose that $L(He)\leq
L(H)-1$. Choose $F\in M^{\prime }(G)$. Since $(v,v^{\prime })\notin F$ we
have $\{(u_{2},v),(u_{3},v)\}\cap F\neq \varnothing $, therefore $%
\{e,(u_{5},u_{6}),(u_{7},u_{8})\}\subseteq F$, hence

\begin{center}
$L(G)=3+L(He)$ (lemma 5).
\end{center}

Choose $F_{1}^{\prime }\in M^{\prime }(H)$, and define $F^{\prime }$ as
follows:

\begin{center}
$F^{\prime }\equiv F_{1}^{\prime }\cup
\{(u_{1},u_{2}),(u_{3},u_{4}),(u_{5},u_{6}),(u_{7},u_{8}),(v,v^{\prime })\}$.
\end{center}

Note that $F^{\prime }\in M(G)$ and

\begin{center}
$\beta (G\backslash F^{\prime })=2+L(H)\geq 3+L(He)=L(G)$, hence

$F^{\prime }\in M^{\prime }(G)$ and $(v,v^{\prime })\in F^{\prime }$,
\end{center}

which is impossible.

Hence $L(H)=L(He)$ and $L(G)=3+L(He)=3+L(H)$. Let us show that $\alpha
(G\backslash \{u_{1},...,u_{4}\})=\beta (G\backslash \{u_{1},...,u_{4}\})$.
Note that

\begin{center}
$8+\beta (H)+L(H)=8+\beta (H)+L(He)=\beta (G)+L(G)=\lambda (G)=8+\lambda
(H)\geq 8+\beta (H)+L(H)$,
\end{center}

hence

\begin{center}
$\lambda (H)=\beta (H)+L(H)$ or $\beta (H)=\alpha (H)$.
\end{center}

(*) and (**) imply that

\begin{center}
$\alpha (G\backslash \{u_{1},...,u_{4}\})=3+\alpha (H)$,

$\beta (G\backslash \{u_{1},...,u_{4}\})=3+\beta (H)$,
\end{center}

we imply that $\alpha (G\backslash \{u_{1},...,u_{4}\})=\beta (G\backslash
\{u_{1},...,u_{4}\})$, and therefore due to hypothesis of induction $%
G\backslash \{u_{1},...,u_{4}\}$ can be built from $K_{1}$ or $K_{2}$ by A,
B, C or D. Note that $G$ is obtained from $G\backslash \{u_{1},...,u_{4}\}$
by operation C since the vertex $v$ is applicable for it. This shows that $G$
can also be built from $K_{1}$ or $K_{2}$ by A, B, C or D.

Case 2: $r=1$. Let $v^{\prime \prime },u_{1},...,u_{6}$ be vertices shown in
the figure below:

\begin{center}
\FRAME{ftbpF}{5.348in}{3.0018in}{0in}{}{}{figure 17.jpg}{\raisebox{-3.0018in}{\includegraphics[height=3.0018in]{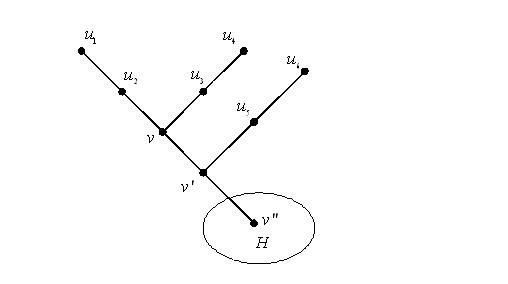}}}

Figure 17\bigskip
\end{center}

Assume:

\begin{center}
$H\equiv G\backslash \{v,v^{\prime },u_{1},...,u_{6}\}$, $e\equiv (v^{\prime
},v^{\prime \prime })$.
\end{center}

We need to consider two cases:

Case 2a: there is no $F\in M^{\prime }(G)$ with $(v,v^{\prime })\in F$.

First of all note that since there is a maximum matching of \ $G$ which does
not contain the edge $(v,v^{\prime })$, we have

\begin{center}
$\beta (G)=4+\beta (H)=3+\beta (He)$, therefore

$\beta (He)=\beta (H)+1$ and $L(He)\leq L(H)$.
\end{center}

Let us show that $\ L(He)=L(H)$. Suppose that $L(He)\leq L(H)-1$. Choose $%
F\in M^{\prime }(G)$. Since $(v,v^{\prime })\notin F$ we have $%
\{(u_{2},v),(u_{3},v)\}\cap F\neq \varnothing $, therefore $%
\{(u_{5},u_{6}),e\}\subseteq F$, hence

\begin{center}
$L(G)=3+L(He)$ (lemma 5).
\end{center}

Choose $F_{1}^{\prime }\in M^{\prime }(H)$, and define $F^{\prime }$ as
follows:

\begin{center}
$F^{\prime }\equiv F_{1}^{\prime }\cup
\{(u_{1},u_{2}),(u_{3},u_{4}),(u_{5},u_{6}),(v,v^{\prime })\}$.
\end{center}

Note that $F^{\prime }\in M(G)$ and

\begin{center}
$\beta (G\backslash F^{\prime })=2+L(H)\geq 3+L(He)=L(G)$, hence

$F^{\prime }\in M^{\prime }(G)$ and $(v,v^{\prime })\in F^{\prime }$,
\end{center}

which is impossible.

Hence $L(H)=L(He)$ and $L(G)=3+L(He)=3+L(H)$. Let us show that $\alpha
(G\backslash \{v,u_{1},...,u_{4}\})=\beta (G\backslash
\{v,u_{1},...,u_{4}\}) $. Note that lemma 2 implies that

\begin{center}
$6+\beta (He)+L(He)=7+\beta (H)+L(H)=\beta (G)+L(G)=\lambda (G)=6+\lambda
(He)\geq 6+\beta (He)+L(He)$,
\end{center}

hence

\begin{center}
$\lambda (He)=\beta (He)+L(He)$ or $\beta (He)=\alpha (He)$.
\end{center}

As (see operation B, lemma 4)

\begin{center}
$\alpha (G\backslash \{v,u_{1},...,u_{4}\})=1+\alpha (He)$,

$\beta (G\backslash \{v,u_{1},...,u_{4}\})=1+\beta (He)$,
\end{center}

we imply that $\alpha (G\backslash \{v,u_{1},...,u_{4}\})=\beta (G\backslash
\{v,u_{1},...,u_{4}\})$, and therefore due to hypothesis of induction $%
G\backslash \{v,u_{1},...,u_{4}\}$ can be built from $K_{1}$ or $K_{2}$ by
A, B, C or D. Note that $G$ is obtained from $G\backslash
\{v,u_{1},...,u_{4}\}$ by operation D since the vertex $v$ is applicable for
it. This shows that $G$ can also be built from $K_{1}$ or $K_{2}$ by A, B, C
or D.

Case 2b: there is $F\in M^{\prime }(G)$ with $(v,v^{\prime })\in F$.

Clearly,

\begin{center}
$\beta (G)=4+\beta (H)$ and, due to lemma 5, $L(G)=2+L(H)$.
\end{center}

Let us show that $\lambda (He)=\lambda (H)$. Lemma 2 implies that

\begin{center}
$6+\beta (H)+L(H)=\beta (G)+L(G)=\lambda (G)=6+\lambda (He)\geq 6+\lambda
(H)\geq 6+\beta (H)+L(H)$,
\end{center}

therefore $\lambda (He)=\lambda (H)=\beta (H)+L(H)$. (*) and (**) imply that

\begin{center}
$\alpha (G\backslash \{u_{1},...,u_{4}\})=2+\alpha (H)$,

$\beta (G\backslash \{u_{1},...,u_{4}\})=2+\beta (H)$,
\end{center}

therefore $\alpha (G\backslash \{u_{1},...,u_{4}\})=\beta (G\backslash
\{u_{1},...,u_{4}\})$, and due to hypothesis of induction $G\backslash
\{u_{1},...,u_{4}\}$ can be built from $K_{1}$ or $K_{2}$ by A, B, C or D.
Note that $G$ is obtained from $G\backslash \{u_{1},...,u_{4}\}$ by
operation C since the vertex $v$ is applicable for it. This shows that $G$
can also be built from $K_{1}$ or $K_{2}$ by A, B, C or D.

The proof of the Theorem is completed.

\bigskip

\textbf{Acknowledgement}. \ We would like to thank Hasmik Sargsyan for her
simplification of the operation B. We are also indebted to Vahe Musoyan and
Anush Tserunyan for their careful reading of the manuscript and for their
useful comments and suggestions that helped us to improve the paper.

\bigskip

\begin{center}
\textbf{References}\bigskip
\end{center}

[1] R. Diestel, Graph theory, Springer-Verlag Heidelberg, New York, 1997,
2000, 2005.

[2] F. Harary, Graph Theory, Addison-Wesley, Reading, MA, 1969.

[3] F. Harary, M.D. Plummer, On the core of a graph, Proc. London Math. Soc.
17 (1967), pp. 305--314.

[4] I. Holyer, The $NP$-completeness of edge coloring, SIAM J. Comput. 10, N
4, 718-720, 1981.

[5] R.R. Kamalian, V. V. Mkrtchyan, On complexity of special maximum
matchings constructing, Discrete Mathematics, to appear.

[6] R.R. Kamalian, V. V. Mkrtchyan, Two polynomial algorithms for special
maximum matching constructing in trees, manuscript.

[7] L. Lovasz, M.D. Plummer, Matching theory, Ann. Discrete Math. 29 (1986).

[8] V. V. Mkrtchyan, On trees with a maximum proper partial 0-1 coloring
containing a maximum matching, Discrete Mathematics 306, (2006), pp. 456-459.

[9] V. V. Mkrtchyan, A note on minimal matching covered graphs, Discrete
Mathematics 306, (2006), pp. 452-455.

[10] V. V. Mkrtchyan, V. L. Musoyan, A. V. Tserunyan, On edge-disjoint pairs
of matchings, Discrete Mathematics 2006, (submitted)

[11] A. V. Tserunyan, Characterization of a class of graphs related to pairs
of disjoint matchings, Discrete Mathematics 2006, (submitted)

[12] D. B. West, Introduction to Graph Theory, Prentice-Hall, Englewood
Cliffs, 1996.

\end{document}